\documentclass[conference]{IEEEtran}
\IEEEoverridecommandlockouts

\usepackage{cite}
\usepackage{booktabs}
\usepackage{amsmath,amssymb,amsfonts}
\usepackage{algorithmic}
\usepackage{graphicx}
\usepackage{textcomp}
\usepackage{xcolor}
\usepackage{graphicx}
\usepackage{multirow}
\usepackage{bbding} 
\def\BibTeX{{\rm B\kern-.05em{\sc i\kern-.025em b}\kern-.08em
    T\kern-.1667em\lower.7ex\hbox{E}\kern-.125emX}}
\begin{document}

\title{Hierarchical Feature Integration for Multi-Signal Automatic Modulation Recognition\\
}

\author{Yunpeng Qu$^{1}$, Yazhou Sun$^{1}$, Bingyu Hui$^{1}$, and Jian Wang$^{1}$\\
\textit{Department of Electronic Engineering, BNRist, Tsinghua University}\\
emails: \{qyp21, sunyz22, hby23\}@mails.tsinghua.edu.cn\\
jian-wang@tsinghua.edu.cn
}

\maketitle

\begin{abstract}
Automatic modulation recognition (AMR) is a crucial step in wireless communication systems, which identifies the modulation scheme from detected signals to provide key information for further processing.
However, previous work has mainly focused on the identification of a single signal, overlooking the phenomenon of multiple signal superposition in practical channels and the signal detection procedures that must be conducted beforehand.
Considering the susceptibility of radio frequency (RF) signals to noise interference and significant spectral variations, we propose a novel Hierarchical Feature Integration (HIFI)-YOLO framework for multi-signal joint detection and modulation recognition.
Our HIFI-YOLO framework, with its unique design of hierarchical feature integration, effectively enhances the representation capability of features in different modules, thereby improving detection performance.
We construct a large-scale AMR dataset specifically tailored for scenarios of the coexistence or overlapping of multiple signals transmitted through channels with realistic propagation conditions, consisting of diverse digital and analog modulation schemes.
Extensive experiments on our dataset demonstrate the excellent performance of HIFI-YOLO in multi-signal detection and modulation recognition as a joint approach.
\end{abstract}


\section{Introduction}
With the rapid proliferation of mobile devices and advancements in wireless communication technologies, there is a growing demand for analyzing and monitoring radio frequency (RF) signals \cite{li2019survey}. 
The ability to detect and recognize RF signals is crucial in the field of cognitive radio, providing support in dynamic spectrum sensing, channel interference monitoring, spectrum sharing, and various applications \cite{stoyanova2020survey}.

Automatic modulation recognition (AMR) plays a critical role in wireless signal recognition, with the primary goal of identifying the modulation scheme of received signals in complex environments \cite{zhou2020deep}. 
Traditional AMR methods primarily relied on signal processing and extracting manually designed features that distinguish signals from noise or signals with different modulations \cite{dobre2007survey}. 
However, designing manual features requires strong prior expert knowledge and lacks generalization when communication systems and environments change \cite{peng2021survey}.
In recent years, deep learning (DL) methods have gradually been introduced into the AMR field and have achieved results far superior to traditional methods.
Structures based on Convolutional Neural Networks (CNNs) \cite{o2016convolutional}, long short-term memory (LSTM) \cite{hong2017automatic}, and transformer \cite{qu2024enhancing}, among others, have shown outstanding performance and strong generalization in AMR due to their representation power of underlying features.
DL methods have become the mainstream solutions in both academic research and industry.

Nevertheless, contemporary research paradigms for AMR are constrained by certain limitations as they tend to idealize the channel conditions excessively.
For instance, existing methodologies often concentrate on the identification of single signals \cite{xing2024joint}, neglecting the potential temporal and spectral overlaps that might manifest in real-world spectrum sensing environments.
Furthermore, current AMR methods typically emphasize signal classification, omitting the detection process of locating signals from the spectrum \cite{vagollari2021joint}, which serves as a fundamental prerequisite for modulation recognition tasks in practical communication systems \cite{prasad2020downscaled}.

Establishing a unified DL framework for joint detection and recognition of multiple signals is an effective solution to address the requirements of spectrum sensing.
To achieve temporal and spectral localization of RF signals, it is imperative to convert signals into spectrograms as a time-frequency representation \cite{uvaydov2024stitching}, allowing us to reframe the multi-signal AMR issue as a classical object detection problem.
RF signals possess unique characteristics compared to natural images, where signals are often subject to significant noise interference \cite{zhang2021novel}, and modulation features may concurrently manifest in both high-frequency (\textit{e.g.}, transient pulses) and low-frequency (\textit{e.g.}, continuous waveforms) spectral variations.
These attributes render the extraction and comprehensive representation of modulation-related features arduous, posing challenges for detection and AMR in complex environments.

\begin{figure*}[t]
\centering
\includegraphics[width=0.9\linewidth]{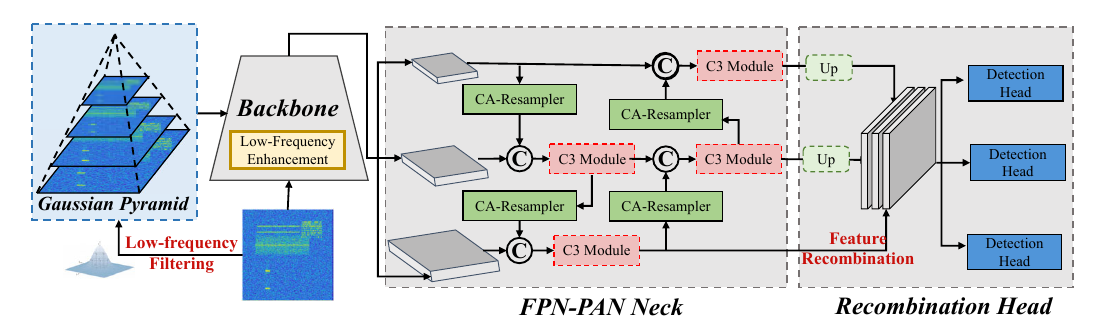}
\caption{Framework diagram of our proposed \textbf{HI}erarchical \textbf{F}eature \textbf{I}ntegration (HIFI)-YOLO. RF signals are converted into a spectrogram as the input to the framework, sequentially passing through the backbone, FPN-PAN neck, and finally obtaining the position and modulation through the predictive head.}
\label{framework}
\end{figure*}
To overcome the above limitation, we believe that the hierarchical integration of features in the time-frequency representation serves as an effective method, which aids in integrating details (\textit{e.g.}, spectral discontinuities) and overall patterns to create a more comprehensive representation and suppress noise interference.
In this paper, we propose a novel \textbf{HI}erarchical \textbf{F}eature \textbf{I}ntegration (HIFI)-YOLO framework for multi-signal joint detection and modulation recognition.
Our HIFI-YOLO framework leverages the classical object detection model YOLO \cite{redmon2016you} as its foundation.
We specifically design multiple stages in our framework to integrate hierarchical features. 
In the backbone, we propose a low-frequency enhancement (LFE) module to aid in signal localization and detection. 
Additionally, we introduce a content-adaptive (CA)-Resampler module in the neck and a recombination head to facilitate the fusion of multi-scale features.
Our contributions are as follows:
\begin{enumerate}
\item{
We propose a novel framework, HIFI-YOLO, to achieve joint signal detection and AMR for multiple signals. 
HIFI-YOLO enhances feature representation through hierarchical feature integration modules, including a low-frequency enhanced backbone, CA-Resampler, and feature recombination head.
}
\item{
We validate the effectiveness of multi-scale feature integration for signal detection and recognition, enhancing representation range and anti-interference capability.
Furthermore, enhancing low-frequency features improves the spectral localization of signals.
}
\item{
Extensive experiments validate the performance of HIFI-YOLO, surpassing baseline methods in terms of recognition accuracy. Comprehensive ablation studies confirm the effectiveness of various components.
}
\end{enumerate}

\section{System Model}
Following the typical wireless communication system, a single received RF signal $r_i(t)$ can be represented as follows:
\begin{equation}
\label{received signal}
r_i(t) = [s_i(t)*h_i(t,\tau)] e^{-2\pi f_i t},
\end{equation}
where $s(t)$, $h(t,\tau)$, $\tau$, and $f_i$ represent the transmitted signal, the channel impulse response function, channel parameters, and the carrier frequency, respectively. "$*$" denotes the convolution operation in the delay time domain.

In communication systems, the receiver simultaneously receives $K$ independent signals existing in the channel, sampled at a fixed rate $f_s$, and subjected to Hilbert transformation.
\begin{equation}
\label{descrete signal}
r[n] = \sum_{i=1}^K r_i[n] + w_n[n], \ n = 0, 1, ..., N-1.
\end{equation}
where $w_n(t)$ represents the Gaussian noise.

Signals are transmitted based on an orthogonal basis, resulting in received signals $r[t]$ having both in-phase and quadrature (I/Q) components.
When locating and detecting multiple signals, we aim to determine the time and frequency positions of the RF signals. Therefore, it is necessary to transform the I/Q signals into the time-frequency domain for processing.
We use the short-time Fourier transform (STFT) to convert the signal $r[n]$ into a spectrogram $X \in \mathbb R^{N_t\times N_f}$:
\begin{equation}
X[m, k]=\sum_{n=0}^{N-1} r[m+n] w[n] e^{-j 2 \pi m k / N}.
\end{equation}
where $w[n]$ is the window function. We then take the logarithm of the spectrum $X$ to calculate the energy gain and perform peak normalization to obtain $\hat{X}$, which can be treated as an image input for processing with our recognition framework.

\section{Methods}
Our framework utilizes YOLO \cite{redmon2016you} as the foundation, which is currently one of the most widely used object detection algorithms, achieving end-to-end learning from pixel values to bounding box coordinates and modulation class probabilities.
As a one-stage method for detection and classification, it better aligns with the real-time processing requirements of spectrum sensing in terms of efficiency.
Due to its outstanding performance in various industrial applications and strong generalization compared to other iterative versions, we have chosen YOLOv5 \cite{Jocher_YOLOv5_by_Ultralytics_2020} as the benchmark architecture.

The overall architecture of HIFI-YOLO, as illustrated in Fig. \ref{framework}, can be divided into three parts: the backbone network, the neck network, and the prediction head. These components are respectively utilized for deep feature extraction, multi-scale feature fusion, and regression of prediction results.
We have specifically designed each of these stages to promote hierarchical feature integration, thereby better suppressing noise interference and enhancing the representation capabilities for RF-related features. These will be elaborated on in detail in the following sections.

\subsection{Low-frequency Enhanced Backbone}
RF signals are highly susceptible to high noise interference during transmission. Therefore, effectively distinguishing signals in the presence of intense noise and accurately locating them pose significant challenges in signal detection.
Due to noise in the channel often manifesting as high-frequency components in the spectrogram and appearing as discrete noise points in the image, we believe that enhancing the low-frequency features of the spectrogram and suppressing high-frequency features will help us better locate signals.

Based on the above considerations, we have specifically introduced a low-frequency enhancement (LFE) module in the backbone network. By applying low-frequency Gaussian filtering to the image to suppress noise, we aim to obtain more robust low-frequency features.
Given the spectrum $\hat{X}$ as the initial input $X_0 \in \mathbb R^{1\times H_0\times W_0}$ where we define $H_0=N_t$ and $W_0=N_f$, we apply layer-by-layer Guassian low-pass filtering to generate the Gussian Pyramid $\{X^{(l)}\}_L$:
\begin{equation}
X^{(l)}= Guassian(X^{(l-1)}), \ l = 1, ..., L.
\end{equation}
where $X^{(l)}\in \mathbb R^{1\times \frac{H_{l-1}}{2}\times \frac{W_{l-1}}{2}}$ is the $l$-th level of the pyramid obtained by Gaussian filtering and downsampling.
In the backbone network of YOLO, the feature maps also gradually decrease in size through convolutional downsampling, following a pyramid structure.
Therefore, we set the number of levels $L$ in the Gaussian pyramid to be equal to the number of downsampling blocks in YOLO, where the feature map $X_l \in \mathbb R^{C_l\times H_l\times W_l}$ in the backbone can correspond to the scale of the spectrum in the Gaussian pyramid.

As $X^{(l)}$ represents the enhanced structures in the denoised spectrum, we specifically designed the LFE module in each block of the backbone to incorporate these low-frequency features that aid in signal localization. 
LFE connects the backbone features and low-frequency information in a residual manner, calculated as follows:
\begin{equation}
X_l = Conv(X_{l-1} + Conv(X^{(l-1)})), \ l = 1, ..., L.
\end{equation}
where the output channels in the two convolutional layer are $C_l$ and $C_{L-1}$.
Based on the LFE module, the low-frequency enhanced backbone network gradually integrates multi-scale Gaussian pyramid information, thereby better suppressing noise interference and preserving essential information.
Our experiments also verify the advantage of enhancing low-frequency features over high-frequency features for signal localization, as noise interference is typically represented as high-frequency variations in the spectrums.

\subsection{Content-Adaptive Resampler}
In the YOLO architecture, the fusion and integration of multi-scale features are achieved through the Feature Pyramid Network (FPN) \cite{lin2017feature} and the Path Aggregation Network (PAN) \cite{liu2018path} as the neck. 
Multi-scale information is passed through both top-down and bottom-up pathways to enhance the semantics of shallow features and compensate for the detail loss in deep features due to downsampling steps, enabling bidirectional flow of hierarchical features.

In this process, matching features of different resolutions for scale alignment is crucial, often achieved through simple nearest-neighbor or bilinear interpolation.
However, these fixed-pattern sampling methods overlook the semantics within the feature maps and cannot dynamically integrate multi-scale features.
Building on prior work in effective upsampling modules, including DySample \cite{liu2023learning}, which employs offset generation for point sampling on feature maps, and CARAFE \cite{wang2019carafe}, which utilizes content-aware upsampling kernels through dynamic convolution, we introduce a Content-Adaptive Resampler in HIFI-YOLO. This module leverages content-aware guidance from upper layers for point sampling and extends its applicability to both downsampling and upsampling modules.

\begin{figure}[t]
\centering
\includegraphics[width=0.92\linewidth]{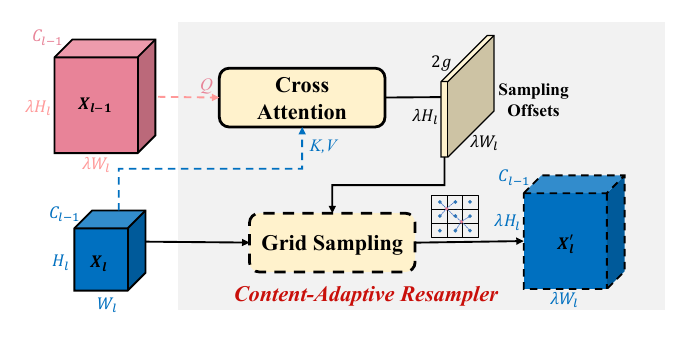}
\caption{Module designs of the Content-Adaptive Resampler. The resampling of features is guided by upper-level features, obtaining the sample offset through a cross-attention mechanism and subsequently performing grid sampling.}
\label{resampler}
\end{figure}
We believe that the resampling process fundamentally aims to achieve alignment and reorganization of features at higher scales. Therefore, the resampling process should be guided by the content of upper-level features, enabling the adaptive selection of semantically related information.
The proposed CA-Resampler is shown in its basic form in Fig. \ref{resampler}. 
Initially, we calculate the semantic correlations between different patches of upper-level features $X_{l-1} \in \mathbb R^{\lambda H_l \times \lambda W_l \times C_{l-1}}$ and lower-level features  $X_{l} \in \mathbb R^{H_l \times W_l \times C_{l-1}}$ in a form similar to cross-attention as the offsets $\mathcal{S} \in \mathbb R^{\lambda H_l \times \lambda H_l \times 2g}$ for the sampling set, where $X_{l}$ has already undergone a convolutional layer to align the number of channels with $X_{l-1}$.
\begin{equation}
 \begin{split}
Q &= X_{l-1}, K= X_l, V= X_lW^v,\\
&\mathcal{S} = Softmax(\frac{QK^T}{\sqrt{C_{l-1}}})V,\\
\end{split}
\end{equation}
where $\mathbf{W}^v \in \mathbb R^{C_{l-1} \times 2g}$ is projection weight, $g$ represents the number of feature partition groups, and $\lambda$ indicates the sampling scale, with options of 2 for upsampling or 0.5 for downsampling. The sampling set $\mathcal{S}$ acts as the sampling offsets in a manner similar to deformable convolution \cite{dai2017deformable}.

Then, we add the content-adaptive offsets $\mathcal{S}$ to the fixed grid points $\mathcal{I}$ of bilinear sampling, then generating the resampled map $X_l^{'}\in \mathbb R^{\lambda H_l \times \lambda H_l \times C_{L-1}}$ through grid sampling:
\begin{equation}
X_l^{'} = GridSample(X_l, \mathcal{S}+\mathcal{I}),
\end{equation}
Our CA-Resampler preserves the resampling spatial structure by maintaining fixed grid points $\mathcal{I}$ and incorporates semantic guidance from the upper scales through generating content-aware offsets, enabling adaptive hierarchical feature perception and integration.
Based on the grid sampling of DySample \cite{liu2023learning}, the lightweight CA-adapter does not introduce excessive computational complexity and extends to both the upsampling and downsampling processes of FPN-PAN.

\subsection{Feature Recombination Head}
YOLO uses multiple independent prediction heads to generate predictions for different scale feature maps $X_l$ in the neck network, corresponding to varying-sized prediction boxes.
We believe that segregating hierarchical predictions may impede modulation recognition, as modulation-related features, including frequency variations and spectral distribution, are present in both the high-frequency and low-frequency parts of the spectrogram, corresponding to features of different scales.

Therefore, in HIFI-YOLO, we propose a feature recombination head to enhance hierarchical feature integration, enabling the utilization of comprehensive information of all the scales for generating prediction results.
Specifically, we uniformly upsample feature maps to match $X_L$, and concatenate them to obtain the recombined $\mathcal{X}$.
$\mathcal{X}$ is processed by independent prediction heads to integrate multi-scale information for generating classification results for objects of different sizes.

In addition, the prediction head of YOLOv5 shares parameters between the classification and localization branches, which is disadvantageous because they handle different types of information.
Following YOLOX \cite{ge2021yolox}, we use decoupled heads to separately predict the position and class probabilities, reducing the number of convolutional layers to one and scaling down the channel size to improve efficiency.
In summary, our feature recombination head is as follows:
\begin{equation}
 \begin{split}
&\mathcal{X} = \mathrm{Concat}([upsample(X_1), ..., X_L]),\\
x&_{reg}^{l}=Head_{reg}^l(\mathcal{X}),\ x_{cls}^{l}=Head_{cls}^l(\mathcal{X}).\\
\end{split}
\end{equation}
where $Head$ is a classifier composed of convolutional layers.

\section{Experiments}
\subsection{Dataset Generation}
\begin{figure}[t]
\centering
\includegraphics[width=0.92\linewidth]{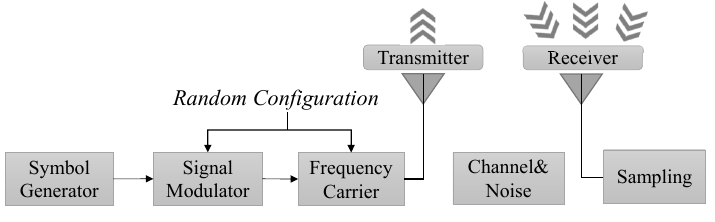}
\caption{Pipeline of RF signal generation.}
\label{fig:dataset}
\end{figure}
\begin{table}[!t]
\caption{Dataset configuration.}
\begin{center}
\label{tab:dataset}
\begin{tabular}{c | c }
    \toprule
\textbf{Parameters} & \textbf{Configuration}\\
    \midrule
& BPSK, QPSK, 8PSK, 16PSK, 16QAM,\\
Modulation type& 64QAM, 2FSK, 4FSK, 8FSK, OOK,\\
& 4ASK, 8ASK, AM-DSB, AM-SSB \\
    \midrule
SNR & -10:2.5:10 (dB)\\
Samples & 36000\\
Signal Length & 204800\\
Sample rate & 200kHz\\
Rician K-factor & 4\\
\midrule
Number of coexisting signals & [3,5]\\
Maximum overlapping range & 0.4 \\
Duration distribution &  'log-uniform', [0.05, 1.0]\\
Bandwidth distribution &  'log-uniform', [0.05, 0.3]\\
\bottomrule
\end{tabular}
\end{center}
\end{table}
To reflect a realistic channel environment, we generate an RF signal dataset with multiple signals coexisting through simulation, supporting the training and evaluation.
The signal generation process is illustrated in Fig. \ref{fig:dataset}. 
Several modulated signals are randomly generated at the transmitter based on pre-designed hyperparameters, occupying different positions in the spectrum. After passing through a complex channel and noise environment, these signals are sampled at the receiver.

Tab. \ref{tab:dataset} summarizes the dataset configuration. Our dataset covers 14 common analog and digital modulation schemes, allowing 3-5 signals to coexist at the receiver with a maximum time-frequency overlap range of 0.4 relative to the signal area.
We simulate a Rician fading channel to replicate a real environment while varying noise levels to assess performance across different signal-to-noise ratios (SNR).
All received signals are jointly displayed on a wideband spectrum at the receiver.
The dataset comprises a total of 36,000 samples.

\subsection{Experimental Setups}
We conduct experiments on HIFI-YOLO using our large-scale multi-signal dataset. 
We divide the datasets for training, validation, and testing by a ratio of 6:2:2.
Our HIFI-YOLO utilizes the lightweight YOLOv5-S as the base model to meet the deployment requirements on mobile devices.
The time-frequency resolution of the input signal's spectrogram is set to $640\times640$, and a 5-level Gaussian pyramid filtering is applied.
The partition groups $g$ in the CA-Resampler are set to 4, while the feature recombination head uses $1\times1$ convolutional kernel to reduce complexity.
The training batch size is set to 32, and the AdamW optimizer is used with a learning rate of 0.002 and weight decay of 0.01.
We use pre-trained YOLO models to accelerate convergence, following other default configurations.

We compare our HIFI-YOLO with several classic object detection methods, including Faster-RCNN \cite{ren2015faster}, SSD \cite{liu2016ssd}, CenterNet \cite{duan2019centernet}, CornetNet \cite{law2018cornernet}, DETR \cite{carion2020end}, and our baseline YOLOv5-S. 
In addition, we compare with YOLO-gIoU \cite{vagollari2021joint}, which is also proposed for RF signal detection.
All experiments are implemented on an NVIDIA GTX 3090 GPU.

\subsection{Comparison with baselines}
\begin{table*}[t]
\centering
\caption{Performance comparison of all methods}
\label{sota_tabel}
\begin{tabular}{ c | cc | cccc }
\toprule
\textbf{Benchmarks} &  \textbf{Params (M)} & \textbf{Flops (G)} & $\mathbf{mAP_{50:95}\ (\%)}$ & $\mathbf{mAP_{50}\ (\%)}$ & \textbf{F1}\\
\midrule
Faster-RCNN \cite{ren2015faster} & 41.4 & 134.3 & 63.1 & 73.8 & 0.702  \\

SSD \cite{liu2016ssd}& 26.3 & 218.2 & 59.3 & 71.3 & 0.673\\

CenterNet \cite{duan2019centernet} & 32.1 & 123.3 & 61.3 & 71.8 & 0.689\\

CornetNet \cite{law2018cornernet} & 201.0 & 1016.1 & 66.2 &  73.2 &  0.709\\ 
DETR \cite{carion2020end} & 41.6 &60.5 & 61.4 & 70.8 & 0.691\\

YOLOv5\cite{Jocher_YOLOv5_by_Ultralytics_2020} &  7.2 & 16.5 & 65.8 &  74.8 &  0.709\\
 
YOLO-gIoU \cite{vagollari2021joint} & 61.5 & 154.7 & 65.5 &  75.1 &  0.707\\
\midrule
\textbf{HIFI-YOLO (Ours)} & 7.7 &  19.1 &  \textbf{68.9} &  \textbf{76.8} &  \textbf{0.730} \\
\midrule
 \textit{Improvement to baseline YOLOv5} &\textit{+6.9\%}&\textit{+15.7\%}&\textit{+4.7\%}& \textit{+2.7\%}&\textit{+3.0\%} \\
\bottomrule
\end{tabular}
\end{table*}
\begin{figure}[t]
\centering
\includegraphics[width=0.9\linewidth]{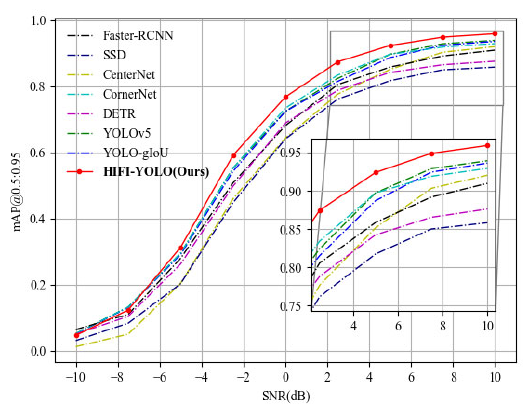}
\caption{Comparison with other methods under different SNR levels.}
\label{fig:sota}
\end{figure}
\begin{figure}[t]
\centering
\includegraphics[width=0.9\linewidth]{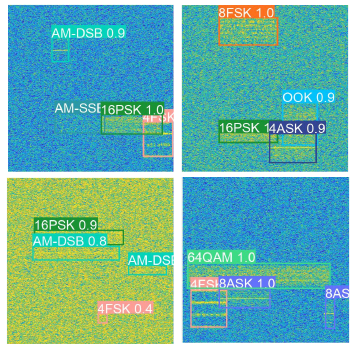}
\caption{Visualization of the detection and modulation recognition results.}
\label{fig:visualization}
\end{figure}

We use the classic mAP and F1 score metrics from the object detection domain as evaluation criteria.
We compare our HIFI-YOLO with other state-of-the-art (SOTA) methods, and the results are illustrated in Tab.~\ref{sota_tabel} and Fig.~\ref{fig:sota}.
HIFI-YOLO outperforms existing SOTA methods across all SNR levels and achieves the highest $\mathrm{mAP_{50:95}}$ (abbreviated as mAP) of 68.9\% and the highest F1 score of 0.730.
In comparison to the original baseline YOLOv5, we achieve a performance improvement of 3\% to 5\% across three metrics by incorporating enhanced hierarchical feature integration into the framework.
Notably, due to the improved noise resistance resulting from the hierarchical feature integration and low-frequency enhancement, HIFI-YOLO demonstrates a significant advantage in low-SNR scenarios. For instance, at -2.5dB, mAP of HIFI-YOLO increases from the original YOLOv5's 55.2\% to 59.1\% (+7.0\%).

Visualization in Fig.~\ref{fig:visualization} demonstrates the ability of HIFI-YOLO to detect, locate, and classify multiple signals that coexist or overlap simultaneously.
HIFI-YOLO can effectively recognize signals in complex scenarios with low SNR (-2.5dB, bottom left) or small signals with short bandwidth.
Even in scenarios where multiple signals overlap complexly in the time-frequency domain, HIFI-YOLO is still able to distinguish and recognize modulation schemes effectively.

We utilize widely-used metrics such as the number of parameters and floating-point operations (FLOPs) to compare the complexity.
Although HIFI-YOLO introduces a slight increase in computational complexity compared to baseline YOLOv5 (mainly coming from the prediction head while other modules are lightweight), it is still much lower than other methods. Moreover, it significantly outperforms in recognition accuracy, showcasing its potential for deployment on mobile devices.

\subsection{Ablation on the LFE module}
\begin{table}[t]
\centering
\caption{Ablation on the Low-frequency enhanced backbone}
\label{tab:lfe}
\begin{tabular}{ cc | ccc }
\toprule
\textbf{Methods} & \textbf{Filtering}  & $\mathbf{mAP_{50:95}(\%)}$ & $\mathbf{mAP_{50}(\%)}$ & \textbf{F1}\\
\midrule
w/o LFE & NA & 67.8 &  75.4 &  0.721 \\
w LFE & Laplacian  & 67.3 &  76.3 &  0.718 \\
w LFE & Guassian & \textbf{68.9} &  \textbf{76.8} &  \textbf{0.730}  \\
\bottomrule
\end{tabular}
\end{table}
We validate the effectiveness of low-frequency feature enhancement. 
We compare two other scenarios: (1) without using the LFE module and (2) using high-pass filtering instead of our low-pass filtering to generate the Laplacian pyramid.

In Tab.~\ref{tab:lfe}, Low-frequency enhancement significantly improves performance compared to not using the LFE module.  
Conversely, high-frequency enhancement with the Laplacian pyramid leads to a decrease in accuracy, supporting our view that enhancing low-frequency information helps suppress noise and accurately locate signals.
\subsection{Ablation on the CA-Resampler}
\begin{table}[t]
\caption{Ablation on the Content-Adaptive Resampler}
\label{tab:resampler}
\centering
\resizebox{0.9\linewidth}{!}{
\begin{tabular}{ cc | ccc }
\toprule
\textbf{Upsampler} & \textbf{Downsampler}  & $\mathbf{mAP_{50:95}(\%)}$ & $\mathbf{mAP_{50}(\%)}$ & \textbf{F1}\\
\midrule
\XSolidBrush & \XSolidBrush & 66.6 &  75.2 &  0.712 \\
\Checkmark& \XSolidBrush  & 67.4 &  75.4 &  0.716 \\
\XSolidBrush& \Checkmark  & 67.8 &  76.3 &  0.726 \\
\Checkmark& \Checkmark & \textbf{68.9} &  \textbf{76.8} &  \textbf{0.730}  \\
\bottomrule
\end{tabular}}
\end{table}
We apply our CA-Resampler for upsampling and downsampling in the Neck stage, and the results are shown in Tab.~\ref{tab:resampler}.
It is evident that applying the CA-Resampler for both upsampler and downsampler improved performance, with a 3.5\% increase in mAP compared to the original sampling module.
This demonstrates that our CA-Resampler, guided by upper-level features, effectively integrates multi-scale features to enhance representational capacity. 
It shows effectiveness in both top-down and bottom-up information propagation, as upper-level features can adaptively perform semantic selection.

\subsection{Ablation on the Recombination Head}
\begin{table}[t]
\centering
\caption{Ablation on the Feature Recombination Head}
\label{tab:head}
\begin{tabular}{ c | ccc }
\toprule
\textbf{Methods}  & $\mathbf{mAP_{50:95}(\%)}$ & $\mathbf{mAP_{50}(\%)}$ & \textbf{F1}\\
\midrule
Original & 67.8 &  76.3 &  0.719 \\
Recombination Head & \textbf{68.9} &  \textbf{76.8} &  \textbf{0.730}  \\
\bottomrule
\end{tabular}
\end{table}
We validate the role of the proposed feature recombination head, which is also the main source of increased computational complexity in HIFI-YOLO (2GFlops increase).
In Tab.~\ref{tab:head}, after applying the recombination head, mAP and F1 scores improved by 1.6\% and 1.5\%.
This validates the effectiveness of hierarchical feature recombination and decoupled predictions for detection and classification, as modulated-related features often manifest at different scales in feature maps.

\section{Conclusion}
In this paper, we proposed HIFI-YOLO, a joint framework integrating signal detection and multi-signal AMR for time-frequency coexisting signals.
HIFI-YOLO was designed for hierarchical feature integration, enhancing the representation capability of multi-scale features.
We built a large-scale dataset with multiple signals coexisting, reflecting real-world conditions.
Experiments showed that HIFI-YOLO achieves outstanding accuracy in identifying overlapping signals.

\section*{Acknowledgment}
This paper is supported by the BNRist projects (No.
BNR20231880004 and No.BNR2024TD03003).

\bibliographystyle{IEEEtran}
\bibliography{ref.bib}

@article{li2019survey,
  title={A survey on deep learning techniques in wireless signal recognition},
  author={Li, Xiaofan and Dong, Fangwei and Zhang, Sha and Guo, Weibin and others},
  journal={Wireless Communications and Mobile Computing},
  volume={2019},
  publisher={Hindawi}
}

@article{stoyanova2020survey,
  title={A survey on the internet of things (IoT) forensics: challenges, approaches, and open issues},
  author={Stoyanova, Maria and Nikoloudakis, Yannis and Panagiotakis, Spyridon and Pallis, Evangelos and Markakis, Evangelos K},
  journal={IEEE Communications Surveys \& Tutorials},
  volume={22},
  number={2},
  pages={1191--1221},
  year={2020},
  publisher={IEEE}
}

@article{zhou2020deep,
  title={Deep learning for modulation recognition: A survey with a demonstration},
  author={Zhou, Ruolin and Liu, Fugang and Gravelle, Christopher W},
  journal={IEEE Access},
  volume={8},
  pages={67366--67376},
  year={2020},
publisher={IEEE}
}

@article{dobre2007survey,
  title={Survey of automatic modulation classification techniques: classical approaches and new trends},
  author={Dobre, Octavia A and Abdi, Ali and Bar-Ness, Yeheskel and Su, Wei},
  journal={IET communications},
  volume={1},
  number={2},
  pages={137--156},
  year={2007},
  publisher={IET}
}

@inproceedings{o2016convolutional,
  title={Convolutional radio modulation recognition networks},
  author={O’Shea, Timothy J and Corgan, Johnathan and Clancy, T Charles},
  booktitle={Engineering Applications of Neural Networks: 17th International Conference, EANN 2016, Aberdeen, UK, September 2-5, 2016, Proceedings 17},
  pages={213--226},
  year={2016},
  organization={Springer}
}

@inproceedings{hong2017automatic,
  title={Automatic modulation classification using recurrent neural networks},
  author={Hong, Dehua and Zhang, Zilong and Xu, Xiaodong},
  booktitle={2017 3rd IEEE International Conference on Computer and Communications (ICCC)},
  pages={695--700},
  year={2017},
  organization={IEEE}
}

@article{qu2024enhancing,
  title={Enhancing automatic modulation recognition through robust global feature extraction},
  author={Qu, Yunpeng and Lu, Zhilin and Zeng, Rui and Wang, Jintao and Wang, Jian},
  journal={IEEE Transactions on Vehicular Technology},
  year={2024},
  publisher={IEEE}
}

@article{xing2024joint,
  title={Joint Signal Detection and Automatic Modulation Classification via Deep Learning},
  author={Xing, Huijun and Zhang, Xuhui and Chang, Shuo and Ren, Jinke and Zhang, Zixun and Xu, Jie and Cui, Shuguang},
  journal={IEEE Transactions on Wireless Communications},
  year={2024},
  publisher={IEEE}
}

@inproceedings{vagollari2021joint,
  title={Joint detection and classification of RF signals using deep learning},
  author={Vagollari, Adela and Schram, Viktoria and Wicke, Wayan and Hirschbeck, Martin and Gerstacker, Wolfgang},
  booktitle={2021 IEEE 93rd Vehicular Technology Conference (VTC2021-Spring)},
  pages={1--7},
  year={2021},
  organization={IEEE}
}

@inproceedings{redmon2016you,
  title={You only look once: Unified, real-time object detection},
  author={Redmon, Joseph and Divvala, Santosh and Girshick, Ross and Farhadi, Ali},
  booktitle={Proceedings of the IEEE conference on computer vision and pattern recognition},
  pages={779--788},
  year={2016}
}

@software{Jocher_YOLOv5_by_Ultralytics_2020,
author = {Jocher, Glenn},
doi = {10.5281/zenodo.3908559},
license = {AGPL-3.0},
month = may,
title = {{YOLOv5 by Ultralytics}},
url = {https://github.com/ultralytics/yolov5},
version = {7.0},
year = {2020}
}

@article{ge2021yolox,
  title={Yolox: Exceeding yolo series in 2021},
  author={Ge, Zheng and Liu, Songtao and Wang, Feng and Li, Zeming and Sun, Jian},
  journal={arXiv preprint arXiv:2107.08430},
  year={2021}
}

@inproceedings{carion2020end,
  title={End-to-end object detection with transformers},
  author={Carion, Nicolas and Massa, Francisco and Synnaeve, Gabriel and Usunier, Nicolas and Kirillov, Alexander and Zagoruyko, Sergey},
  booktitle={European conference on computer vision},
  pages={213--229},
  year={2020},
  organization={Springer}
}

@inproceedings{law2018cornernet,
  title={Cornernet: Detecting objects as paired keypoints},
  author={Law, Hei and Deng, Jia},
  booktitle={Proceedings of the European conference on computer vision (ECCV)},
  pages={734--750},
  year={2018}
}

@inproceedings{duan2019centernet,
  title={Centernet: Keypoint triplets for object detection},
  author={Duan, Kaiwen and Bai, Song and Xie, Lingxi and Qi, Honggang and Huang, Qingming and Tian, Qi},
  booktitle={Proceedings of the IEEE/CVF international conference on computer vision},
  pages={6569--6578},
  year={2019}
}

@inproceedings{liu2016ssd,
  title={Ssd: Single shot multibox detector},
  author={Liu, Wei and Anguelov, Dragomir and Erhan, Dumitru and Szegedy, Christian and Reed, Scott and Fu, Cheng-Yang and Berg, Alexander C},
  booktitle={Proceedings of the European conference on computer vision (ECCV)},
  pages={21--37},
  year={2016},
}

@article{ren2015faster,
  title={Faster r-cnn: Towards real-time object detection with region proposal networks},
  author={Ren, Shaoqing and He, Kaiming and Girshick, Ross and Sun, Jian},
  journal={Advances in neural information processing systems},
  volume={28},
  year={2015}
}

@inproceedings{dai2017deformable,
  title={Deformable convolutional networks},
  author={Dai, Jifeng and Qi, Haozhi and Xiong, Yuwen and Li, Yi and Zhang, Guodong and Hu, Han and Wei, Yichen},
  booktitle={Proceedings of the IEEE international conference on computer vision},
  pages={764--773},
  year={2017}
}

@inproceedings{liu2023learning,
  title={Learning to upsample by learning to sample},
  author={Liu, Wenze and Lu, Hao and Fu, Hongtao and Cao, Zhiguo},
  booktitle={Proceedings of the IEEE/CVF International Conference on Computer Vision},
  pages={6027--6037},
  year={2023}
}

@inproceedings{wang2019carafe,
  title={Carafe: Content-aware reassembly of features},
  author={Wang, Jiaqi and Chen, Kai and Xu, Rui and Liu, Ziwei and Loy, Chen Change and Lin, Dahua},
  booktitle={Proceedings of the IEEE/CVF international conference on computer vision},
  pages={3007--3016},
  year={2019}
}

@inproceedings{liu2018path,
  title={Path aggregation network for instance segmentation},
  author={Liu, Shu and Qi, Lu and Qin, Haifang and Shi, Jianping and Jia, Jiaya},
  booktitle={Proceedings of the IEEE conference on computer vision and pattern recognition},
  pages={8759--8768},
  year={2018}
}

@inproceedings{lin2017feature,
  title={Feature pyramid networks for object detection},
  author={Lin, Tsung-Yi and Doll{\'a}r, Piotr and Girshick, Ross and He, Kaiming and Hariharan, Bharath and Belongie, Serge},
  booktitle={Proceedings of the IEEE conference on computer vision and pattern recognition},
  pages={2117--2125},
  year={2017}
}

@article{prasad2020downscaled,
  title={A downscaled faster-RCNN framework for signal detection and time-frequency localization in wideband RF systems},
  author={Prasad, KNR Surya Vara and D’souza, Kevin B and Bhargava, Vijay K},
  journal={IEEE transactions on wireless communications},
  volume={19},
  number={7},
  pages={4847--4862},
  year={2020},
  publisher={IEEE}
}

@article{peng2021survey,
  title={A survey of modulation classification using deep learning: Signal representation and data preprocessing},
  author={Peng, Shengliang and Sun, Shujun and Yao, Yu-Dong},
  journal={IEEE Transactions on Neural Networks and Learning Systems},
  volume={33},
  number={12},
  pages={7020--7038},
  year={2021},
  publisher={IEEE}
}

@article{zhang2021novel,
  title={A novel automatic modulation classification scheme based on multi-scale networks},
  author={Zhang, Hao and Zhou, Fuhui and Wu, Qihui and Wu, Wei and Hu, Rose Qingyang},
  journal={IEEE Transactions on Cognitive Communications and Networking},
  volume={8},
  number={1},
  pages={97--110},
  year={2021},
  publisher={IEEE}
}

@inproceedings{uvaydov2024stitching,
  title={Stitching the Spectrum: Semantic Spectrum Segmentation with Wideband Signal Stitching},
  author={Uvaydov, Daniel and Zhang, Milin and Robinson, Clifton Paul and D’Oro, Salvatore and Melodia, Tommaso and Restuccia, Francesco},
  booktitle={IEEE INFOCOM 2024-IEEE Conference on Computer Communications},
  pages={2219--2228},
  year={2024},
  organization={IEEE}
}
\end{document}